%% file: main.tex
\documentclass[sigconf, natbib=true]{acmart}

\AtBeginDocument{%
  }

\setcopyright{acmlicensed}
\copyrightyear{2026}
\acmYear{2026}
\acmDOI{XXXXXXX.XXXXXXX}
\acmConference[ArXiv Preprint]{}
\acmISBN{978-1-4503-XXXX-X/2018/06}

\usepackage{multirow}
\usepackage{booktabs}

\begin{document}

\title{A Replicability Study of XTR}

\author{Rohan Jha}
\email{rjha5@jh.edu}
\orcid{TODO}
\affiliation{%
  \institution{Johns Hopkins University}
  \city{Baltimore}
  \state{Maryland}
  \country{USA}
}
\author{Reno Kriz}
\email{rkriz1@jh.edu}
\orcid{TODO}
\affiliation{%
  \institution{Human Language Technology Center of Excellence}
  \city{Baltimore}
  \state{Maryland}
  \country{USA}
}\author{Benjamin Van Durme}
\email{vandurme@jhu.edu}
\orcid{TODO}
\affiliation{%
  \institution{Johns Hopkins University}
  \city{Baltimore}
  \state{Maryland}
  \country{USA}
}

\renewcommand{\shortauthors}{Jha et al.}

\begin{abstract}

\input{sections/00-abstract}
\end{abstract}




\maketitle
\section{Introduction}
\label{sec:intro}

\input{sections/10-introduction}
\section{Background and Preliminaries}
\label{sec:background}
\input{sections/20-background}

\section{Experiment Replication}
\label{sec:replication}
\input{sections/30-results-and-analysis}
\subsection{Experimental Setup}
\label{sec:setup}
\input{sections/31-experimental-setup}

\subsection{Retrieval Results}
\label{sec:results}

\input{sections/32-retrieval-results}

\subsection{Token Retrieval Dynamics}
\label{sec:analysis}
\input{sections/33-token-retrieval-dynamics}

\section{Conclusion}
\label{sec:conclusion}

\input{sections/40-conclusion}

\section*{Acknowledgements}
We acknowledge and thank Antoine Chaffin, Raphaël Sourty, and Luca Arnaboldi for their helpful discussions and support in integrating this work into PyLate.

\bibliographystyle{ACM-Reference-Format}
\bibliography{main}

\end{document}

%% file: sections/00-abstract.tex
The XTR (conteXtual Token Retrieval) algorithm is a modification to ColBERT retrieval that avoids the costly step of fully gathering and reranking the candidates' embeddings by imputing their missing similarity scores from the initial token retrieval step.
The original work proposes a modified training objective as necessary for effective XTR retrieval, arguing that standard ColBERT token scoring is unsuitable for imputation.
In this paper, we replicate both the XTR retrieval algorithm and its modified training objective, and extend the evaluation to knowledge-distillation (KD) training and efficient retrieval engines (PLAID and WARP).
We confirm the token-level matching characteristics claimed in the original work, but fail to replicate XTR's overall effectiveness advantage over ColBERT under a controlled comparison.
We further show that XTR's training modification has a concrete mechanistic consequence for modern retrieval engines: by flattening ColBERT's characteristically peaked token score distribution, XTR training yields more discriminative centroid scores and thus more efficient IVF-based retrieval under PLAID and WARP.
The utility of XTR training is therefore not limited to the low-$k'$ regime originally studied, but extends to any deployment setting where IVF-based engines are used.
These findings offer practitioners concrete guidance on how and when to use XTR as their multi-vector retriever.

%% file: sections/10-introduction.tex
ColBERT \cite{khattab_2020_colbert, santhanam_2022_colbertv2} has established itself as an extremely capable architecture for first-stage retrieval, approximating the expressivity of cross-encoder joint attention \cite{nogueira_2020_passagererankingbert} while maintaining the offline indexability of bi-encoders.
However when compared to single-vector sparse and dense retrievers \cite{formal_2021_splade, karpukhin_2020_dpr}, which can benefit more straightforwardly from the decades of optimizations in lexical retrieval and approximate $k$-nearest-neighbor search \cite{ding_2011_blockmaxwand, bruch_2024_seismic, malkov_2020_hnsw}, late interaction falls behind in terms of efficiency.
As such, there have been many followup works aiming to reduce index size and query latency by shrinking various dimensions of the indexed document embeddings (token vector size, precision, and count \cite{jha_2024_jinacolbertv2, santhanam_2022_colbertv2, macavaney_2025_constbert, clavie_2024_reducing}) and efficiently produce and rerank candidate documents using MaxSim scoring.

One such work, XTR (conteXtual Token Retrieval) \cite{lee_2023_xtr}, makes ColBERT retrieval faster by eliminating the costly step of gathering each candidate document's full set of token embeddings. 
Instead, XTR imputes the missing query-document token scores in the MaxSim matrix as the minimum score that \textit{was} retrieved for each query token in its initial $k'$ retrieved tokens.
This serves as a sensible upper bound on the true score for a document token's score with a given query token, achieved without needing to load any further data from the index.

In addition to the XTR retrieval algorithm, Lee et al. \cite{lee_2023_xtr} propose a modification to the InfoNCE-style contrastive loss used to train ColBERT.
The XTR modification simulates token retrieval at training time by masking query-document token scores $\mathbf{q}_i^\top \mathbf{d}_i$ from contributing to the overall document score if $\mathbf{d}_i$ is not ``retrieved'' among the top-$k_{\text{train}}$ in-batch tokens similar to $\mathbf{q}_i$.
They claim the purpose of this modified objective is to better align the training setting to how the model will be used at inference time by optimizing for token retrieval ability as a prerequisite for document scoring.


While Lee et al. \cite{lee_2023_xtr} released code for XTR retrieval alongside a base-sized XTR model,\footnote{\href{https://huggingface.co/google/xtr-base-en}{\texttt{google/xtr-base-en} on Hugging Face}} no first-party training implementation was ever released.
The XTR retrieval algorithm on its own has already been replicated using the released model, serving as the baseline for WARP \cite{scheerer_2025_warp}, a followup work to XTR that applies PLAID-like optimizations \cite{santhanam_2022_plaid} for even more efficient search.
Subsequent implementations of WARP similarly only use the original model or use ColBERT-trained models for XTR retrieval.
This paper therefore aims to replicate the XTR training objective in order to evaluate the claims and analysis of the original paper, and to further evaluate XTR-trained models under WARP --- closing the loop between the training objective and the retrieval engine designed for it.

First, we validate our implementation of XTR retrieval by using the released model, then train XTR models using various $k_{\text{train}}$ in order to evaluate its overall retrieval effectiveness and its token retrieval characteristics.
Our evaluation experiments in Section \ref{sec:results} examine how XTR training and its parameter $k_{\text{train}}$ affect retrieval effectiveness across various candidate token budgets $k'$ and retrieval engines.
Importantly, we fail to replicate the same significant overall bump in retrieval effectiveness when comparing XTR with ColBERT, indicating that it is not simply a strictly better system over ColBERT as initially suggested.
Our analysis experiments in Section \ref{sec:analysis} confirm several key findings from the original work --- XTR relies less on lexical signals for token retrieval and is more robust with small initial retrieval $k'$ --- and extend them.
We show that XTR training flattens the characteristically peaked token score distribution of ColBERT, which has a concrete consequence for IVF-based retrieval engines like PLAID and WARP where score separability governs candidate-set size and thus query latency.

%% file: sections/20-background.tex



\paragraph{ColBERT Retrieval} 
ColBERT encodes queries and documents independently into sequences of contextualized token embeddings using a bidirectional encoder, producing $\mathbf{Q} = [\mathbf{q}_1, \dots, \mathbf{q}_n] \in \mathbb{R}^{n \times d}$ and $\mathbf{D} = [\mathbf{d}_1, \dots, \mathbf{d}_m] \in \mathbb{R}^{m \times d}$ for a query and document respectively.
Relevance is scored via the MaxSim operator $f_{\text{ColBERT}}$, which for each query token finds the maximum cosine similarity against all document tokens and sums the result:
\begin{equation}
    f_{\text{ColBERT}}(Q, D) = \sum_{i=1}^{n} \max_{j \in [m]} \mathbf{q}_i^\top \mathbf{d}_j
\end{equation}
Because document embeddings are computed offline and indexed ahead of time, ColBERT retains the indexability advantages of bi-encoders while capturing fine-grained token-level interactions that approximate cross-encoder expressivity.
Just as in the dense single-vector setting, scoring the whole corpus $D$ with $f_{\text{ColBERT}}$ would be woefully inefficient.
Thus in practice, ColBERT retrieval first does a top-$k'$ approximate-nearest-neighbor lookup for each of its $n$ query tokens, then loads the union of retrieved document candidates' embeddings and scores them using $f_{\text{ColBERT}}$.
However, computing $f_{\text{ColBERT}}$ for any candidate document still requires loading all $m$ of its token embeddings.
Using $k' = 4,000$, with $n = 32$ query tokens, $m=300$ document tokens, and $d=128$-dimensional half-precision vectors, each query would require gathering and scoring \textit{conservatively}\footnote{Assuming 75\% redundancy both within and across query token retrieval.} 614 MB of embeddings, constituting a key latency bottleneck.

A number of works have addressed the efficiency costs of late interaction by compressing the indexed representations.
ColBERTv2 \cite{jha_2024_jinacolbertv2} introduces residual compression, quantizing token embeddings to a centroid $C$ plus a residual vector with 1--4 bits per dimension.
PLAID \cite{santhanam_2022_plaid} builds on this with a multi-stage pipeline that uses centroid interaction for cheap candidate generation and rough scoring before performing full-fidelity scoring with the residuals on a subset of the candidates.
Both methods significantly reduce the time spent on the initial token retrieval, but still spend a large portion of their time decompressing the residual embeddings in order to fully compute $f_{\text{ColBERT}}$. 

\paragraph{XTR Retrieval}
XTR takes a complementary approach: rather than compressing the index, it modifies the retrieval algorithm to avoid loading all token embeddings for each candidate.
XTR first performs a token-level $k'$-nearest-neighbor search for each query token $\mathbf{q}_i$, retrieving the top-$k'$ most similar tokens from the entire index.
For a candidate document $d$, only tokens that appeared in some query token's top-$k'$ have known scores; the remaining entries are imputed as $m_i$, the minimum score retrieved for $\mathbf{q}_i$, serving as an upper bound on any non-retrieved token's true similarity.
This yields the XTR scoring function $f_{\text{XTR}}$, which approximates $f_{\text{ColBERT}}$ without loading any additional token embeddings beyond those already retrieved.
\begin{equation}
    f_{\text{XTR}}(Q, D) = \sum_{i=1}^n \max_{j \in [m]} \left[ \mathbf{A}_{ij}\mathbf{q}_i^\top \mathbf{d}_j + (1 - \mathbf{A}_{ij})m_i \right]
\end{equation}
Where $\mathbf{A}_{ij}$ is defined to be 1 when $\mathbf{d}_j$ is in the top-$k'$ retrieved token embeddings for $\mathbf{q}_i$, else 0. 
Notably, this formulation leaves open the option to impute the missing score $m_i$ with some other cheap-to-compute function.
We review the original authors' experiments with other imputation methods in Section \ref{sec:results}.
With some modification, the core of the PLAID algorithm --- IVF-based candidate generation and compressed residual encoding --- is adapted to XTR's more lightweight retrieval in the form of WARP \cite{scheerer_2025_warp}.

\paragraph{ColBERT Training}
ColBERT models are trained similarly to single-vector retrievers, simply replacing the cosine similarity scoring function with $f_{\text{ColBERT}}$ and doing contrastive learning \cite{khattab_2020_colbert, karpukhin_2020_dpr, oord_2019_contrastivepredictivecoding} on query/document pairs with in-batch and mined hard negatives, or knowledge-distillation \cite{santhanam_2022_colbertv2} with a KL objective to match a teacher's (usually a more expensive cross-encoder reranker) scores for a positive and set of mined negatives.
Subsequent work has also explored better aligning ColBERT training with its quantized state during inference by jointly training the encoder and PQ index \cite{fang_2022_jmpq}.

\paragraph{XTR Training} 
In addition to the retrieval algorithm $f_{\text{XTR}}$, Lee et al. \cite{lee_2023_xtr} propose a modified training objective $f_{\text{XTR}_{\text{train}}}$ that simulates token retrieval at training time.
\begin{align}
    f_{\text{XTR}_{\text{train}}}(Q, D ; k_{\text{train}}) &= \frac{1}{Z} \sum_{i=1}^n \max_{j \in [m]} \left[ \mathbf{A}^{\text{train}}_{ij}\mathbf{q}_i^\top \mathbf{d}_j \right] \\
    Z &= \sum_{i=1}^n \max_{j \in [m]} \mathbf{A}^{\text{train}}_{ij}  
\end{align}
For each query token $\mathbf{q}_i$, a document token $\mathbf{d}_j$ contributes to its document's score only if it ranks among the top-$k_{\text{train}}$ most similar document tokens to $\mathbf{q}_i$ across the batch ($\mathbf{A}^{\text{train}}_{ij} = 1$).
Otherwise --- unlike at inference time where it would be imputed --- the score is dropped.
It is also normalized by $Z$, the number of query tokens in $Q$ that ``retrieve'' at least one document token in $D$.
We confirm $Z$ normalization to be a necessary component to achieve training stability, most evident during early training when few or no tokens of $D$ are retrieved.

%% file: sections/30-results-and-analysis.tex
We begin by replicating the core retrieval results of XTR by examining the relationship between initial token retrieval budget $k'$ and downstream document retrieval effectiveness.
These findings transition us to a discussion of the token retrieval-level dynamics that XTR's training modification induces, including replicating the originally claimed trends of improved retrieved token relevance with lower dependence on lexical matching.

We describe our experimental setup, with details on the the replication of both retrieval and training components in the next section.
Our experiments best fit under ACM's definition of replicability as they are conducted by a different team with a different setup.

%% file: sections/31-experimental-setup.tex
Our experiments use two configurations, which we designate as configurations (\textbf{I}) and (\textbf{II}), corresponding to our original replication experiments and subsequent modernizing extensions (KD training, PLAID/WARP indexing).
We describe the common and differing features of both configurations and make clear in subsequent sections which configuration is in use by including (\textbf{I}) or (\textbf{II}) in the figure caption or discussion.

\paragraph{Models and Code}
We implement all components of XTR within PyLate \cite{chaffin_2025_pylate}, a library built on top of Sentence Transformers \cite{reimers_2019_sentencebert} meant for training and serving ColBERT models.
We also port the only released original XTR model to the PyLate wrapper format in order to validate its original effectiveness.
The implementation of the XTR training objective is based on an open-source implementation.\footnote{\href{https://github.com/primeqa/primeqa/tree/xtr}{\texttt{https://github.com/primeqa/primeqa/tree/xtr}}} \cite{bhat_2025_ibm_xtr_workshop} 
However, we identify and correct a bug in the implementation that causes the token masking simulating retrieval not to occur, and another that failed to normalize embeddings during training, which had a temperature-like effect on training dynamics.
We integrate XTR training and retrieval into PyLate,\footnote{\href{https://github.com/lightonai/pylate}{\texttt{https://github.com/lightonai/pylate}}} and we release all models used in this paper.\footnote{\href{https://huggingface.co/collections/robro612/xtr-replicability}{\texttt{robro612/xtr-replicability}}}

\paragraph{Training (\textbf{I})}
We finetune our models from \texttt{bge-small-en-v1.5}\footnote{\href{https://huggingface.co/BAAI/bge-small-en-v1.5}{\texttt{BAAI/bge-small-en-v1.5}}} \cite{shitao_2023_bge_embedding}, a single-vector encoder.
While the original ColBERT and XTR models trained directly from a pretrained encoder like BERT \cite{devlin_2019_bert}, it is now common to initialize multi-vector models from a well-trained single-vector model \cite{jha_2024_jinacolbertv2, chaffin_2025_pylate}.
Models are trained with a contrastive loss with in-batch negatives for 50,000 steps with a batch size of 196 and learning rate of $3 \times 10^{-5}$ on a triplet version of the 64-way MS MARCO training hard negatives that ColBERTv2 used.\footnote{\href{https://huggingface.co/datasets/bclavie/msmarco-10m-triplets}{\texttt{bclavie/msmarco-10m-triplets}}}
The only difference between XTR and ColBERT is how the scores for the loss function are computed. ColBERT use $f_{\text{ColBERT}}$, while XTR uses $f_{\text{XTR}_{\text{train}}}$ with $k_\text{train} \in \{64, 128, 256, 512\}$ and clamps $Z$ to $10^{-3}$ when it would otherwise be zero. 
When not specified in the experimental results, we use a $k_\text{train}$ of 128.
Following \cite{lee_2023_xtr}, we omit the \texttt{[Q]} and \texttt{[D]} prefix marker tokens used in previous ColBERT models \cite{khattab_2020_colbert}.

\paragraph{Training (\textbf{II})}
We again train contrastively using the same data as (\textbf{I}) with the same number of steps and learning rate, but from \footnote{\href{https://huggingface.co/answerdotai/ModernBERT-base}{\texttt{ModernBERT-base}}} -- a pretrained bidirectional language model with no prior retrieval tuning -- with a batch size of 64.
From these contrastive checkpoints, we continue fine tuning with a KL-based distillation loss for 20,000 steps with a batch size of 16 and 15 negatives per example on the same reranker-scored MS MARCO data\footnote{\href{https://huggingface.co/datasets/lightonai/ms-marco-en-bge-gemma}{\texttt{lightonai/ms-marco-en-bge-gemma}}} that trained \texttt{GTE-ModernColBERT-v1},\footnote{\href{https://huggingface.co/lightonai/GTE-ModernColBERT-v1}{\texttt{lightonai/GTE-ModernColBERT-v1}}} a state-of-the-art ColBERT model.
Following \cite{clavie_2024_jacolbertv2.5}, during distillation, we apply min-max normalization to student and teacher scores which normalizes the score ranges to $[0,1]$.
Following \cite{chaffin_2026_colbertzero} we use a learnable temperature parameter, initialized at 0.2 and 0.05 respectively for the contrastive ColBERT and XTR models (based on early experiments) and 1.0 for the distillation models.
We hypothesize that a lower temperature (less normalization) is required for contrastive XTR due to the greater sparsity of its learning signal due to the masking in $f_{\text{XTR}_{\text{train}}}$.
Based on the poor results of XTR with $k_{\text{train}}=64$ in (\textbf{I})'s experiments, we omit it from (\textbf{II}) and instead train $\text{XTR}_{\text{multi}}$, which stacks equally weighted XTR loss functions with $k_{\text{train}} \in \{128,256,512\}$. 
Note that we also omit $\text{XTR}_{128}$, as it consistently diverged during training.
We suspect that $k_{\text{train}}=128$ is too sparse for the larger batch of $16 \times 16$ documents with hard negatives.

\paragraph{Encoding} 
Following the defaults listed in prior ColBERT work and the PyLate library, we encode documents up to 300 tokens, and encode queries with 32 tokens (48 for SciFact, SCIDOCS, and TREC-COVID, 64 for ArguAna) including the standard \texttt{[MASK]} padding augmentation.
We load models and encode queries and documents in FP16 precision.

\paragraph{Indexing and Retrieval (\textbf{I})}
All experiments using configuration (\textbf{I}) do XTR retrieval as described by the original work with a ScaNN \cite{sun_2023_scann_1_soar, guo_2020_scann_2_avq} index for token retrieval.
ScaNN indexing follows the original work’s configuration, with tree-based partitioning into 2,000 leaves, of which 200 are searched per token query. 
The index also employs single-codebook (global) anisotropic quantization with a quantization threshold of 0.1.
In addition, for ColBERT retrieval which requires the ability to recover the document embeddings, we store a flattened version of the documents' embeddings in full precision alongside the index, which are not organized beyond the document level.
The token retrieval parameter $k'$ will be specified for individual experiments, however when not specified, we use $k' = 4,000$ and $40,000$ for ColBERT and XTR, respectively, reflecting the original work's configuration.

\paragraph{Indexing and Retrieval (\textbf{II})}
Experiments for the more modern configuration use PLAID and WARP retrieval engines for efficiency and to validate our findings in a more realistic deployment setting.
We use the FastPlaid \cite{sourty_2025_fastplaid} and xtr-warp-rs\cite{motserrat_2025_warp_rs} implementations of PLAID and WARP, respectively \cite{sourty_2025_fastplaid, motserrat_2025_warp_rs} for core retrieval results, but use ScaNN for token retrieval analysis.
Both implementations support GPU-accelerated indexing and search, and we thus evaluate both with 1 NVIDIA H100 GPU with 80GB of VRAM.
For PLAID, we use 2-bit residuals and leave the rest of the parameters as their robust PyLate defaults.
For WARP, we found the default parameters to be too conservative, and thus increase its $n_{\text{probes}}$ to 32 and cluster size threshold $t'$ (related to XTR's missing score imputation) to 100,000.
Given that PLAID only uses $n_{\text{probes}}=8$, an important determiner of overall retrieval quality and query latency, we justify setting WARP's higher as a correction of overly-conservative defaults, noting that even with $n_{\text{probes}}=32$, WARP far surpasses PLAID's efficiency, discussed in Section \ref{sec:results}.


\paragraph{Datasets and Evaluation}
Across our experiments we evaluate on a subset of the BEIR benchmark \cite{thakur_2021_beir} which span a diverse range of domains and corpus sizes (AR: ArguAna. FQ: FiQA-2018. MS: MS MARCO. NF: NFCorpus. NQ: Natural Questions. QU: Quora. SD: SCIDOCS. SF: SciFact. TC: TREC-COVID. TO: Touché-2020 v2.) and on the LoTTE benchmark \cite{santhanam_2022_colbertv2} (LI: Lifestyle. WR: Writing. RE: Recreation. TE: Technology. SC: Science. PO: Pooled.), using the search queries from the test split.
We evaluate with the standard metrics for each benchmark: nDCG@10 and Recall@100 for BEIR, MRR@10 for MS MARCO, and Success@5 for LoTTE.
Unfortunately, evaluating specifically our ColBERT models with PLAID on MS MARCO, our largest dataset, proved to be computationally prohibitive. 
We discuss why this is the case for our ColBERT but not identically-sized XTR models in Section \ref{sec:analysis}.

%% file: sections/32-retrieval-results.tex
\input{tables/00-beir-main}

Table \ref{tab:results} presents the results of XTR and ColBERT retrieval on BEIR datasets using experimental configuration (\textbf{I}).
In the first section of the table, we present our reproduction of the original work's results using their model in our indexing and retrieval framework. 
The per-dataset results largely agree with the original works results, while being closer to Scheerer et al.'s \cite{scheerer_2025_warp} independent and lower reproduction of the original results.
The latter half of the table presents the results of our own trained XTR and ColBERT models.
We observe --- counter to the original findings --- that ColBERT maintains an edge over XTR (1.3 points over the best XTR model), and in fact maintains most ($>99\%$) of its effectiveness when retrieving with $f_{\text{XTR}}$ as well.
This would seem to imply that ColBERT-trained models are already set to take advantage of the efficiency gains of XTR's retrieval algorithm, but this table constitutes just one operating point on the Pareto graph (where $k'$ is a surrogate for query latency).
Figure \ref{fig:robust_to_k} tells a more nuanced story.

\input{figures/00-robust-to-k/robust_to_kprime}

Figure \ref{fig:robust_to_k} shows that XTR training induces a greater robustness to small $k'$.
That is, XTR's retrieval quality with small $k'$ (more efficient) appears to be a function of its training parameter, where a higher $k_{\text{train}}$ tends to permit greater effectiveness (both nDCG@10 and Recall@100) when retrieving more candidates, but at the cost of a sharper decline when retrieving fewer.
Conversely, we note that $f_{\text{ColBERT}}$ is even more capable of preserving its effectiveness while retrieving very few tokens per query token, as this does not make the MaxSim scoring matrix sparser as it does with $f_{\text{XTR}}$.
This points to XTR's training objective not being universally useful.

\input{tables/01-imputations}

Table \ref{tab:results} and Figure \ref{fig:robust_to_k} corroborate that in the original setting of $k'=40,000$, an off-the-shelf ColBERT trained model is adequate to approximate $f_{\text{ColBERT}}$ retrieval effectiveness with the efficiency gains of $f_{\text{XTR}}$.
XTR's training objective finds its utility in the low-$k'$ regime, which is motivated by cases in which token retrieval is a significant fraction of the total scoring time, such as with very large datasets or slow indices.



Noting that XTR manages to preserve recall well in the small-$k'$ setting, we ask whether it might be possible to compose $f_{\text{XTR}}$ and $f_{\text{ColBERT}}$ into $f_{\text{XTR}\to \text{ColBERT}}$ by doing full ColBERT reranking on a smaller subset of $k''$ documents.
However we find that subsequent ColBERT reranking yields negligible improvement (0.0 points nDCG@10, 0.1 point Recall@100) when reranking up to $k''=16,000$ candidate documents.\footnote{50\% of the theoretical maximum candidates with $|\mathbf{Q}| =32$ and $k'=1000$}


Another method that might improve XTR's retrieval effectiveness without resorting to loading document token embeddings is modifying the imputation heuristic for estimating missing scores $m_i$.
We replicate the original work's exploration of different imputation methods in Table \ref{tab:imputations}, concluding similarly that min-imputation is sufficient.
The original work finds a marginal benefit to power-law imputation, however we fail to replicate this marginal improvement.
We speculate that the marginal benefit of power-law imputation is sensitive to both k' and the extrapolation factor (i.e., the multiplier determining how far beyond k' the missing score $m_i$ is estimated), due to the variable shape of token score distributions.

\input{tables/02-config2-results}

\begin{figure}[htp!]
    \centering
    \includegraphics[width=1.0\linewidth]{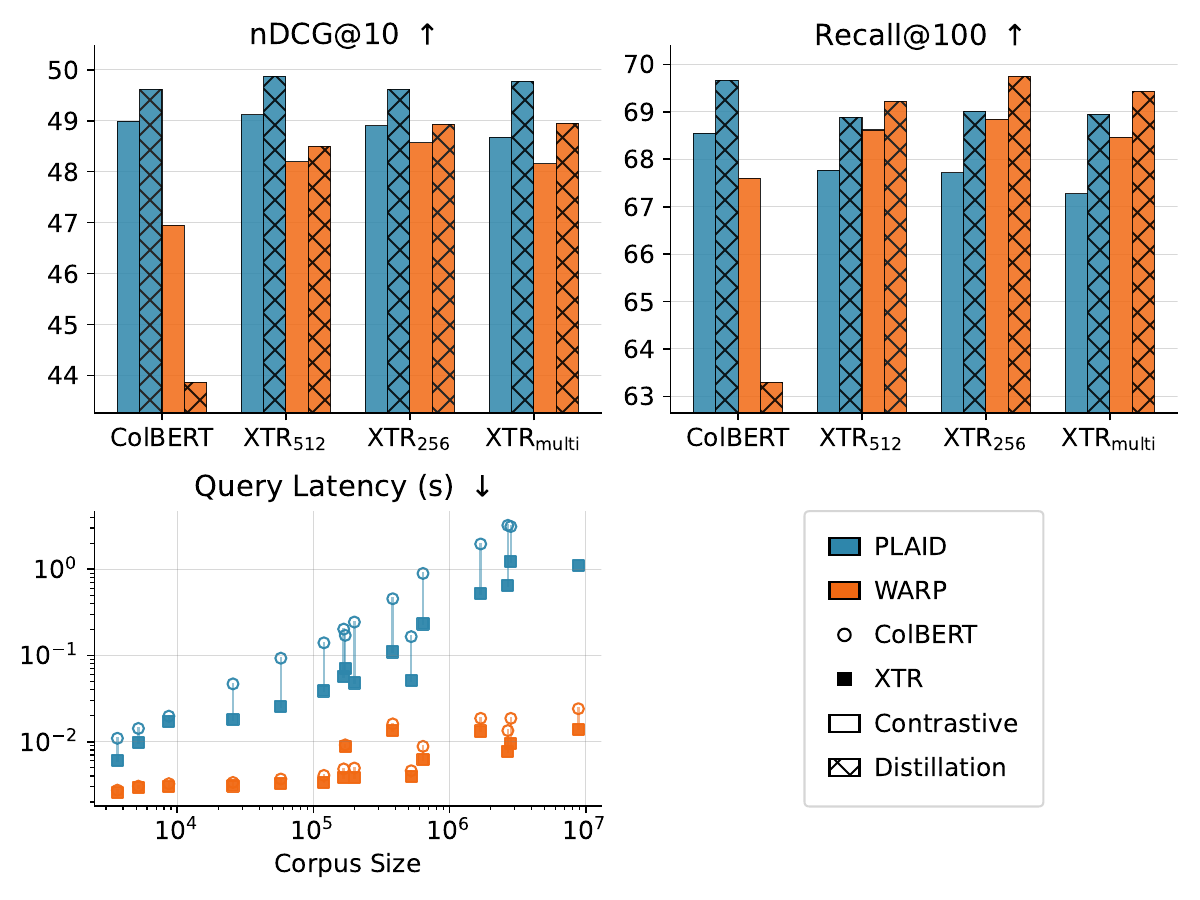}
    \vspace{-2em}
    \caption{Mean nDCG@10 and Recall@100 (top) and per-dataset query latency (bottom) for ColBERT and XTR models on both PLAID and WARP indices. Using configuration (\textbf{II}).}
    \label{fig:retrieval_bars_and_latency_scatter}
    \vspace{-1em}
\end{figure}

Moving beyond naïve XTR retrieval using a ScaNN index and contrastively trained XTR model, we examine XTR in a more contemporary setting: with models trained with distillation losses using multi-vector-specific centroid-based approximate retrieval engines like PLAID and WARP. 
Table \ref{tab:config2_mainresults} presents the results of retrieval on BEIR and LoTTE datasets using experimental configuration (\textbf{II}).
Along with Figure \ref{fig:retrieval_bars_and_latency_scatter} (top), this reconfirms the gap between ColBERT- and XTR-style retrieval with PLAID and WARP, although XTR-trained models are capable of achieving higher \textit{recall} with WARP than PLAID.
The best model using WARP lags underperforms best model using PLAID by 0.6 points on BEIR and 1.0 point on LoTTE.

Notably, however, the best model for both collections using PLAID is, narrowly, an XTR model. 
This indicates that XTR training does not necessarily degrade its ability to be served with ColBERT-style retrieval, and in fact there is evidence, discussed in Section \ref{sec:analysis}, that it may confer beneficial token-level scoring properties.
ColBERT models, on the other hand, demonstrate a marked drop in both nDCG@10 and Recall@100 when served with WARP, contrary to the finding in Figure \ref{fig:robust_to_k} that it is capable (with sufficiently high $k'$) of retrieving with $f_{\text{XTR}}$.
ColBERT models' poorer WARP effectiveness is further widened by continued KL-based distillation training, which confers a benefit to its PLAID retrieval quality.

Finally, we observe in Figure \ref{fig:retrieval_bars_and_latency_scatter} --- as expected --- that WARP is considerably faster than PLAID.
Over all experiments, WARP achieves a geometric mean speedup of $16.1\times$ more queries per second (QPS -- inverse of latency) over PLAID.
When splitting on model type, however, we further discover that XTR models get a mean speedup of $15.1\times$ when switching from PLAID to WARP, while ColBERT gets $29.5\times$ but is still slower with WARP.
This is a consequence of the consistently higher latency ColBERT has compared to XTR models \textit{even when both are using PLAID (and to a lesser extent, WARP)}.
We discuss the cause of this phenomenon in the next section.

%% file: tables/00-beir-main.tex
\begin{table*}[htp!]
    \centering
    \begin{tabular}{llcccccccccc}
        \toprule
        Model & Scoring ($f$) & AR & FQ & NF & NQ & QU & SD & SF & TC & TO & Avg. \\
        \midrule
        \multirow{3}{*}{XTR (Lee et al. \cite{lee_2023_xtr})} & XTR (Lee et al. \cite{lee_2023_xtr}) & 40.7 & 34.7 & 34.0 & 53.0 & 86.1 & 14.5 & 71.0 & 73.6 & 31.3 & 48.8 \\
        & XTR (Scheerer et al. \cite{scheerer_2025_warp}) & -- & 34.1 & 33.5 & -- & 86.0 & 14.3 & 69.6 & -- & 31.2 & N/A \\
        & XTR (Ours) & 41.0 & 34.2 & 33.4 & 50.5 & 85.6 & 14.3 & 69.1 & 73.3 & 32.3 & 48.2 \\
        \midrule
        $\mathrm{XTR}_{64}$ & XTR & 37.9 & 33.4 & 33.1 & 48.2 & 86.7 & 16.5 & 62.0 & 76.0 & 23.5 & 46.4 \\
        $\mathrm{XTR}_{128}$ & XTR & 38.2 & 34.4 & 33.8 & 49.8 & 87.0 & 17.0 & 65.1 & 78.2 & 23.8 & 47.5 \\
        $\mathrm{XTR}_{256}$ & XTR & 39.8 & 35.2 & 33.8 & 50.7 & 86.9 & 16.8 & 65.2 & 77.0 & \underline{24.6} & 47.8 \\
        $\mathrm{XTR}_{512}$ & XTR & \underline{40.2} & 36.1 & 34.4 & 51.6 & 86.9 & 17.2 & 66.9 & 74.6 & \textbf{25.2} & 48.1 \\
        \multirow{2}{*}{ColBERT} & XTR & \textbf{41.4} & \underline{36.8} & \textbf{34.4} & \underline{52.8} & \textbf{87.5} & \underline{18.6} & \underline{70.4} & \underline{78.4} & 23.6 & \textbf{49.3} \\
        & ColBERT & \textbf{41.4} & \textbf{37.2} & \underline{34.5} & \textbf{53.1} & \textbf{87.5} & \textbf{18.6} & \textbf{70.6} & \textbf{78.5} & 23.4 & \textbf{49.4} \\
        \bottomrule
    \end{tabular}
    \vspace{0.5em}
    \caption{nDCG@10 on BEIR datasets, using experimental configuration (\textbf{I}). 
    Per the original work, ColBERT retrieval uses $k' = 4{,}000$ and XTR uses $k' = 40{,}000$.}
    \label{tab:results}
    \vspace{-1.5em}
\end{table*}

%% file: figures/00-robust-to-k/robust_to_kprime.tex
\begin{figure}
    \centering
    \vspace{1em} 
    \includegraphics[width=\columnwidth]{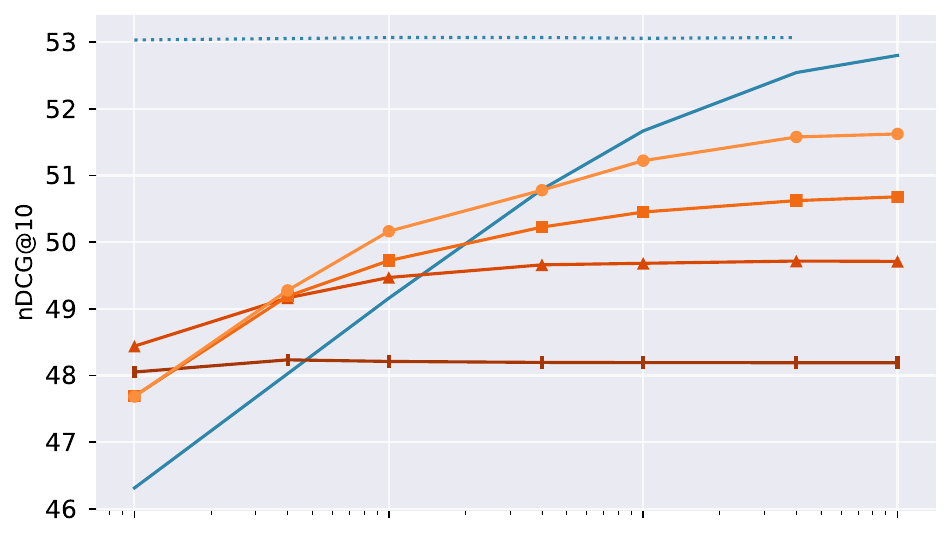}
    \includegraphics[width=\columnwidth]{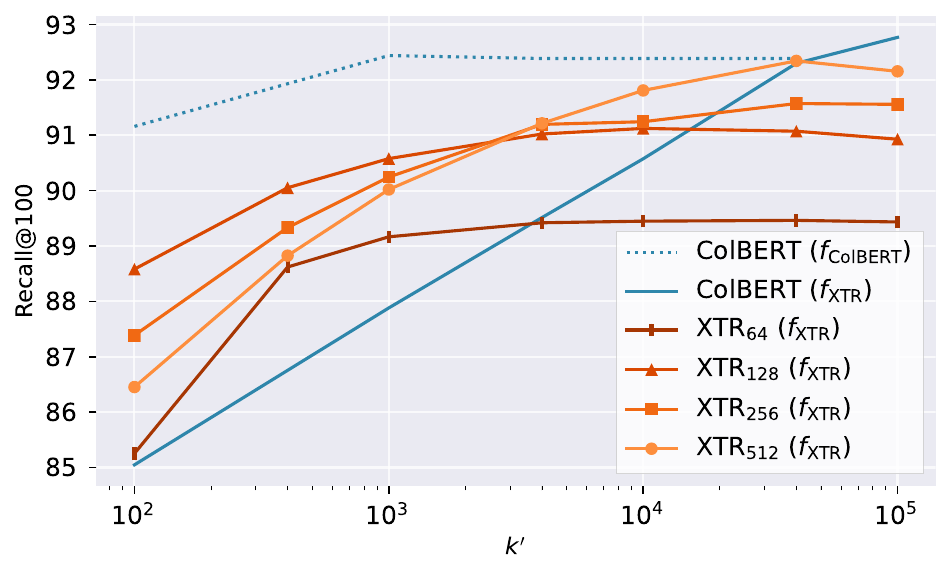}
    \vspace{-2em}
    \caption{
        nDCG@10 (top) and Recall@100 (bottom) vs number of tokens $k'$ initially retrieved per query token for ColBERT and $\text{XTR}_{k_{\text{train}}}$ models on Natural Questions (NQ) using configuration (\textbf{I}).
        $f_{\text{ColBERT}}$ and $f_{\text{XTR}}$ indicate retrieval using ColBERT or XTR scoring. 
    }
    \label{fig:robust_to_k}
    \vspace{-1em}
\end{figure}

%% file: tables/01-imputations.tex
\begin{table}[htp!]
    \centering
    \begin{tabular}{lccccc}
    \toprule
        Dataset & Mean & P10 & Min & Pow & Zero \\
        \midrule
        FiQA & \textbf{34.3} & 34.2 & 34.2 & 34.2 & 33.8 \\
        NFCorpus & 33.2 & 33.1 & \textbf{33.5} & 33.0 & 32.9 \\
        Natural Questions & 49.7 & 49.7 & \textbf{49.8} & 49.6 & 48.9 \\
        TREC-COVID & 77.3 & 77.7 & \textbf{78.2} & 78.0 & 77.4 \\
        \bottomrule
    \end{tabular}
    \vspace{0.5em}
    \caption{Comparing $m_i$ imputation methods for $f_{\text{XTR}}$ in order of optimistic $\to$ pessimistic estimate of $m_i$. Mean: impute $m_i$ as the mean score retrieved for $\mathbf{q}_i$. P10: use bottom 10th percentile of retrieved. Min: normal XTR, use minimum retrieved score. Pow: fit a power law $p$ to the top-$k'$ scores then estimate $m_i$ as $p(100k')$. Zero: $m_i = 0$.}
    \label{tab:imputations}
    \vspace{-1.5em}
\end{table}

%% file: tables/02-config2-results.tex
\begin{table*}[htp!]
\centering
\small
\setlength{\tabcolsep}{4pt}
\begin{tabular}{lll c cccccccccc ccccccc}
\toprule
& & & \textbf{MS} & \multicolumn{10}{c}{\textbf{BEIR}} & \multicolumn{7}{c}{\textbf{LoTTE}} \\
\cmidrule(lr){4-4} \cmidrule(lr){5-14} \cmidrule(lr){15-21}
Index & Stage & Model & MS & AR & FQ & NF & NQ & QU & SD & SF & TC & TO & Avg. & LI & WR & RE & TE & SC & PO & Avg. \\
\midrule
\multirow{8}{*}{\rotatebox[origin=c]{90}{WARP}} & \multirow{4}{*}{\rotatebox[origin=c]{90}{Con.}} 
& ColBERT & 32.4 & 29.9 & 34.2 & 30.8 & 47.7 & 82.4 & 14.9 & 63.7 & 75.2 & 27.5 & 45.1 & 85.0 & 78.8 & 70.4 & 65.6 & 58.7 & 69.9 & 71.4 \\
& & XTR$_{512}$ & 34.1 & 31.3 & 34.9 & 31.8 & 49.3 & 85.2 & 15.0 & 69.5 & 75.2 & 26.1 & 46.5 & 84.1 & 78.7 & 71.4 & 65.9 & 56.1 & 70.8 & 71.2 \\
& & XTR$_{256}$ & 33.6 & 32.7 & 35.2 & 31.2 & 50.0 & 85.4 & 14.6 & 69.1 & 76.7 & 29.2 & 47.1 & 83.5 & 81.1 & 71.8 & 66.3 & 58.7 & 71.5 & 72.1 \\
& & XTR$_{\text{multi}}$ & 34.2 & 32.3 & 34.9 & 31.0 & 49.2 & 85.6 & 15.0 & 66.8 & 76.5 & 29.1 & 46.7 & 83.5 & 80.0 & 71.1 & 66.3 & 58.0 & 71.2 & 71.7 \\
\cmidrule{2-21}
& \multirow{4}{*}{\rotatebox[origin=c]{90}{Dist.}} 
& ColBERT & 31.2 & 17.0 & 12.9 & 33.6 & 48.2 & 81.6 & 12.9 & 52.5 & 72.4 & 30.5 & 40.2 & 83.2 & 81.8 & 68.4 & 62.8 & 55.6 & 69.9 & 70.3 \\
& & XTR$_{512}$ & 33.7 & 29.7 & 35.7 & 34.2 & 51.7 & 84.8 & 15.8 & 70.5 & 74.3 & 28.7 & 47.3 & 83.4 & 80.5 & 71.5 & 64.3 & 58.0 & 71.8 & 71.6 \\
& & XTR$_{256}$ & 33.8 & 32.8 & 35.9 & 33.0 & 51.4 & 85.6 & 15.8 & 71.3 & 78.1 & 30.0 & 48.2 & 84.4 & 81.1 & 69.9 & 65.6 & 60.3 & 70.6 & 72.0 \\
& & XTR$_{\text{multi}}$ & 33.9 & 32.8 & 35.6 & 32.7 & 52.3 & 86.0 & 15.7 & 69.8 & 74.9 & \textbf{31.9} & 48.0 & 84.4 & 82.1 & 72.1 & 64.6 & 58.5 & 71.5 & 72.2 \\
\midrule
\multirow{8}{*}{\rotatebox[origin=c]{90}{PLAID}} & \multirow{4}{*}{\rotatebox[origin=c]{90}{Con.}} 
& ColBERT & - & 33.5 & 37.1 & 32.7 & 52.7 & 85.0 & 14.9 & 70.5 & 78.3 & 25.7 & 47.8 & 85.9 & 81.2 & 69.4 & 67.3 & 58.8 & 71.2 & 72.3 \\
& & XTR$_{512}$ & 35.3 & 32.4 & 36.1 & 32.7 & 52.1 & 85.9 & 14.6 & 69.7 & \textbf{80.2} & 26.5 & 47.8 & 84.3 & 79.2 & \textbf{72.3} & 66.6 & 58.8 & 70.5 & 72.0 \\
& & XTR$_{256}$ & 34.9 & 32.7 & 36.1 & 32.2 & 52.1 & 86.2 & 14.1 & 68.5 & 79.2 & 26.5 & 47.5 & 84.3 & 80.8 & 72.2 & 66.8 & \textbf{60.8} & 69.9 & 72.5 \\
& & XTR$_{\text{multi}}$ & 34.3 & 33.0 & 36.3 & 31.7 & 51.7 & 86.5 & 14.6 & 68.1 & 76.2 & 27.6 & 47.3 & 84.3 & 81.1 & 71.5 & \textbf{68.3} & 58.5 & 69.8 & 72.2 \\
\cmidrule{2-21}
& \multirow{4}{*}{\rotatebox[origin=c]{90}{Dist.}} 
& ColBERT & - & \textbf{33.9} & 36.9 & 34.4 & \textbf{54.7} & 84.4 & 16.2 & \textbf{72.2} & 77.0 & 27.3 & 48.6 & 84.9 & \textbf{83.3} & 71.4 & 64.9 & 58.0 & 71.3 & 72.3 \\
& & XTR$_{512}$ & \textbf{36.1} & 32.5 & \textbf{37.4} & \textbf{34.8} & 54.6 & 85.9 & \textbf{16.4} & 70.8 & 77.6 & 29.3 & \textbf{48.8} & \textbf{86.5} & 82.3 & 71.8 & 67.0 & 60.0 & \textbf{71.9} & \textbf{73.2} \\
& & XTR$_{256}$ & 35.2 & 33.4 & 36.8 & 33.5 & 53.7 & 86.8 & 15.9 & 70.6 & 77.9 & 28.5 & 48.6 & 85.6 & 81.9 & 72.2 & 67.0 & 58.0 & 70.9 & 72.6 \\
& & XTR$_{\text{multi}}$ & 35.0 & 33.6 & 36.7 & 33.7 & 54.6 & \textbf{86.9} & 16.0 & 69.7 & 76.1 & 30.9 & 48.7 & 86.2 & 82.5 & 71.4 & 66.8 & 59.2 & 71.7 & 73.0 \\
\bottomrule
\end{tabular}
\caption{MRR@10, nDCG@10, and Success@5 on MS MARCO, BEIR, and LoTTE, respectively for XTR and ColBERT models at both training stages (contrastive, distillation) and on both retrieval engines (PLAID, WARP), using configuration (\textbf{II}). Missing MS MARCO scores for ColBERT using PLAID are explained in Section \ref{sec:analysis}.}
\label{tab:config2_mainresults}
\vspace{-2em}
\end{table*}

%% file: sections/33-token-retrieval-dynamics.tex
\input{figures/02-token-score-distributions/score_distribution_nfcorpus}

Figure \ref{fig:score_distribution} plots the distribution of the top-1000 token retrieval scores on all NFCorpus queries for both contrastive and distilled models in configuration (\textbf{II}).
Our findings replicate the curious phenomenon in the original work where ColBERT exhibits a spike of many document tokens having extremely high relevance scores,\footnote{We do not have the original authors' ColBERT model to independently verify} regardless of their actual relevance to the input query.
This phenomenon is further amplified for ColBERT by continued distillation training, whereas it slightly widens XTR's score distribution.
However, we note that other KD-trained ColBERT models, e.g. the original ColBERTv2 or the state-of-the-art ModernColBERT-v1, exhibit different distributions.
$\text{XTR}_{k_{\text{train}}}$ variants share a similar distribution to the original work's model, distinct from ColBERT.

While the original work relates $f_{\text{ColBERT}}$ score distributions to a failure to propagate adequate training signal, we extend this analysis to search efficiency.
ColBERT's tight, peaked score distribution provides an explanation for the large gap in query latency between ColBERT and XTR models under identical dataset and index settings, represented by the bar between points in Figure \ref{fig:retrieval_bars_and_latency_scatter} (bottom).
Concretely, PLAID and WARP's efficiency depends on each query token's centroid scores being discriminative: for each of the $N_q$ query tokens, the top-$n_{\text{probes}}$ centroids are selected and their corresponding document ids are unioned to form the candidate set. 
When the score distribution is narrow, the top-$n_{\text{probes}}$ centroids are less distinguished from the remaining irrelevant centroids.
This has two reinforcing effects: different query tokens select less overlapping centroid sets (approaching $N_q \times n_{\text{cells}}$ unique centroids rather than a smaller semantically coherent set), and each selected centroid covers a broader, less semantically focused slice of the corpus. 
The resulting candidate set is substantially larger --- on FiQA with $n_{\text{probe}}=8$, ColBERT produces 15,172 candidates per query (26.3\% of the corpus) versus 10,323 for $\text{XTR}_{256}$ (17.9\%) --- and the downstream centroid pruning threshold (which is also harder to set with a narrow score distribution) cannot compensate, as the operations it gates (loading per-document centroid codes, tensor materialization) scale with the candidate count, not the number of centroids surviving pruning.
Figure \ref{fig:retrieval_bars_and_latency_scatter} (bottom) shows that this is a material problem that affects PLAID (and WARP to a lesser extent), even when the expensive decompression stage is more strictly gated.

Embedding anisotropy provides a useful but incomplete lens for understanding why these score distributions arise.
When token embeddings concentrate in a narrow subspace of the available unit sphere --- as measured on NFCorpus documents by mean pairwise cosine similarity (0.732 for ColBERT vs.\ 0.517 for $\text{XTR}_{256}$) or effective dimensionality (1.8 vs.\ 3.7) --- query-centroid scores are compressed into a narrow band, directly producing the non-discriminative scores described above. 
However, anisotropy as a scalar measure is not fully predictive: ModernColBERT-v1 is the most anisotropic model we evaluate (effective dimensionality 1.2) yet achieves 31.7 QPS on FiQA compared to 13--15 QPS for ColBERT on PLAID, closer to $\text{XTR}_{256}$'s 36.7 QPS and consistent with a score distribution that retains a workable bimodal structure for effective centroid discrimination despite high overall anisotropy. 
The operative property is thus not isotropy per se, but whether the centroid score distribution admits a separable gap between relevant and irrelevant centroids.

\begin{figure}
    \centering
    \includegraphics[width=\linewidth]{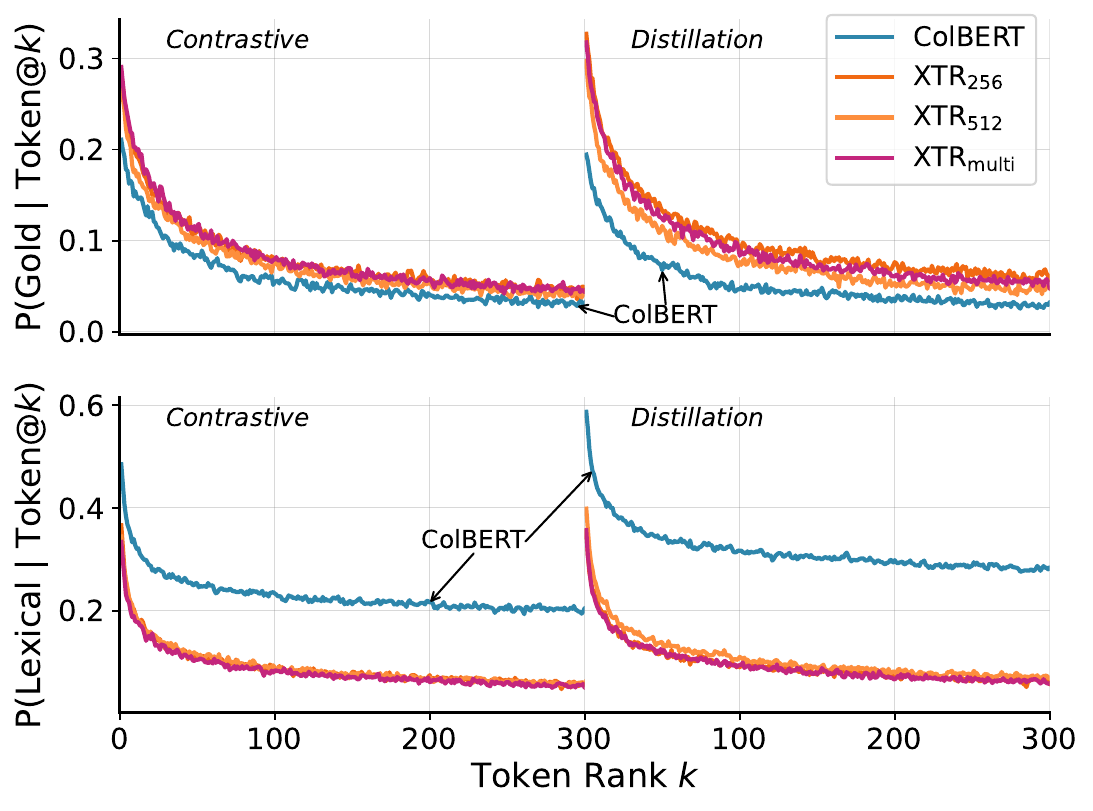}
    \vspace{-2em}
    \caption{Token retrieval characteristics of ColBERT and $\text{XTR}_{k_{\text{train}}}$ on NFCorpus using configuration (\textbf{II}). We plot the empirical probability of document tokens retrieved at each rank being from a gold relevant document (top) or being a lexical match with its query token (bottom).} 
    \label{fig:pgold_plexical_paired}
    \vspace{-1.25em}
\end{figure}


While the score distribution explains the efficiency gap, it characterizes embedding geometry rather than retrieval content. 
To understand how XTR training supports effective retrieval at low $k'$ and under WARP, we replicate the original work's examination of the relevance and lexical characteristics of the tokens that seed candidate generation.

Figure \ref{fig:pgold_plexical_paired} validates the original work's dual findings that the XTR training objective 1) increases the occurrence of relevant documents at every retrieval depth (top) and decreases the likelihood that a given retrieved token will be a lexical match for its query compared to ColBERT.
Comparing contrastive (left) and distillation (right), we see that distillation training increases both gold candidate token retrieval and lexical match probability, albeit more for ColBERT than XTR.
We observe that the described behavioral differences are made more pronounced by extreme values of $k_{\text{train}} \to 0$ and conversely are diminished as $k_{\text{train}} \to \infty$ and thus $f_{\text{XTR}_{k_{\text{train}}}} \to f_{\text{ColBERT}}$.
The original work argues that as the matching signal at training time becomes more sparse --- mimicking the fewer real token scores in XTR retrieval compared to ColBERT --- the model learns to better contextualize its token representations.
This means that XTR token representations 1) disambiguate term senses more finely and 2) bridge vocabulary mismatch better, two canonical challenges that learned dense term embeddings aim to address. 
While this apparently superior token retrieval behavior is not sufficient to make XTR-style retrieval overcome its inherently less exact subsequent document scoring (Figure \ref{fig:robust_to_k}), it does seem to also improve XTR-trained models ability for ColBERT-style retrieval (Table \ref{tab:config2_mainresults}); better candidate retrieval \textit{does} improve overall retrieval in this regime.

%% file: figures/02-token-score-distributions/score_distribution_nfcorpus.tex
\begin{figure}[h!]
    \centering
    \includegraphics[width=\linewidth]{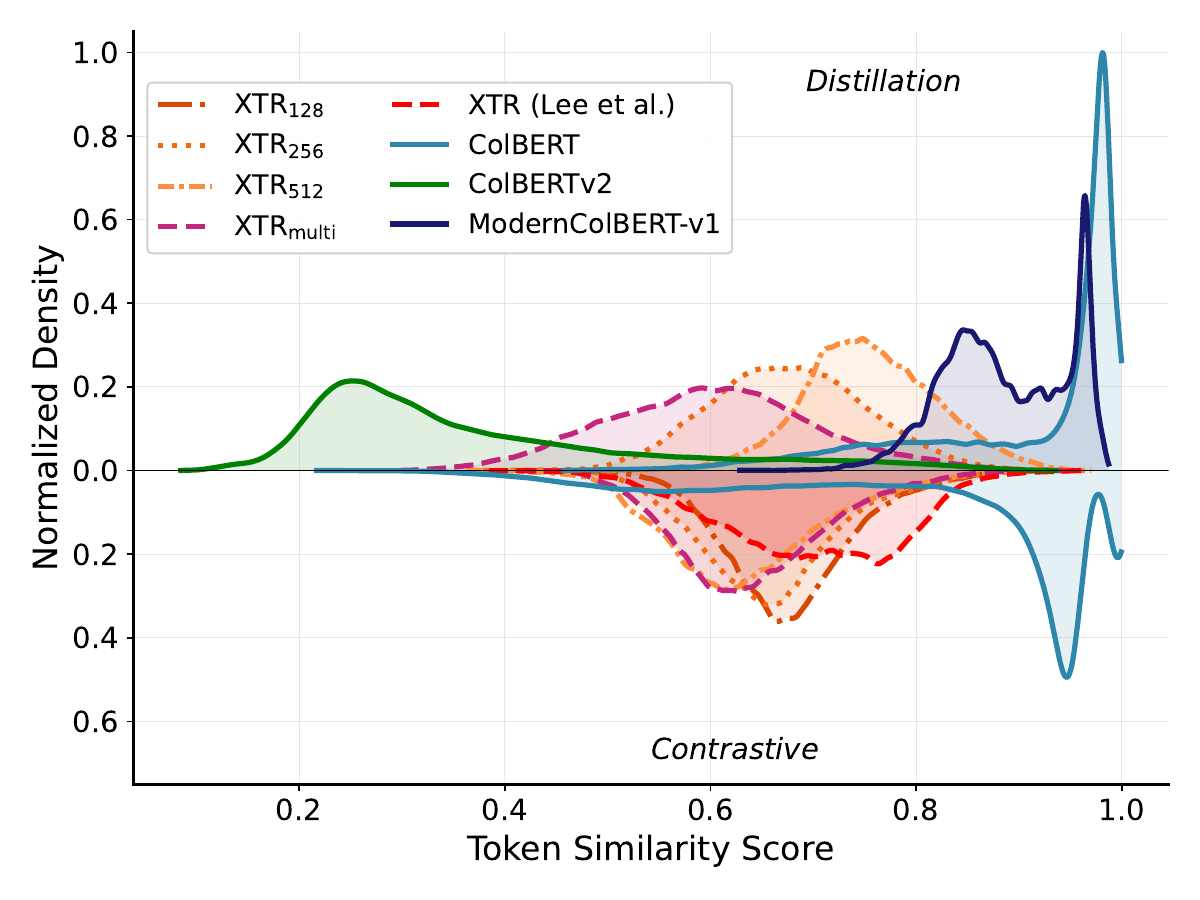}
    \vspace{-2em}
    \caption{Token retrieval score distributions for FiQA (FQ). Distillation training sharpens ColBERT scores towards 1 and widens XTR score ranges.}
    \label{fig:score_distribution}
    \vspace{-0.5em}
\end{figure}

%% file: sections/40-conclusion.tex
In this work, we successfully replicate both the XTR retrieval algorithm and its associated training objective modification, then extend its evaluation to a modern setting --- training a distilled XTR model, evaluating with the WARP engine.
Three findings emerge.

First, XTR's purported effectiveness advantage over ColBERT does not replicate under a controlled comparison.
Retrieval with $f_{\text{XTR}}$ does not surpass $f_{\text{ColBERT}}$ on nDCG@10.
Neither alternative imputation methods nor $f_{\text{XTR}\to\text{ColBERT}}$ reranking pipelines bridge this gap --- suggesting the bottleneck lies not in the imputation heuristic itself but in how token-level scores aggregate into document scores under the XTR framework.

Second, XTR's modified training objective nevertheless produces models with measurably improved token retrieval characteristics: more consistent retrieval of relevant document tokens across depths, and reduced dependence on lexical matching.
These improvements are attributable to the $k_{\text{train}}$ parameter, which simulates token-level retrieval during training by masking non-retrieved scores.

Third, we show that these token-level behaviors have a concrete mechanistic consequence for modern retrieval engines.
ColBERT's tight, peaked score distribution yields non-discriminative centroid scores under PLAID- and WARP-style IVF retrieval, inflating the candidate set and thus query latency.
XTR training flattens this distribution away from 1, producing more efficient retrieval.
This reframes the practitioner tradeoff: XTR training is \textit{more} necessary for WARP than for the naïve XTR retrieval setup originally studied, and under PLAID it still confers marginal retrieval quality gains over ColBERT while also improving efficiency.
The utility of XTR training is thus not limited to the low-$k'$ regime, but extends to any deployment setting where IVF-based engines are used.

Future work might improve the XTR training objective or scoring algorithm to further narrow the retrieval quality gap with ColBERT.
One promising direction is adaptive retrieval strategies that dynamically allocate and schedule the token retrieval budget between full MaxSim and XTR imputation scoring, permitting a more efficient tradeoff between quality and latency, particularly for single-core retrieval in production.